\documentclass[onecolumn,floatfix,showpacs,showkeys,preprintnumbers,nofootinbib,superscriptaddress,11pt]{revtex4}
\usepackage[utf8]{inputenc}
\usepackage[sort&compress]{natbib}
\usepackage{ulem}
\usepackage{bm}
\usepackage{times}
\usepackage{amssymb,amsbsy,amsmath,amsfonts}
\usepackage{graphicx}
\usepackage{float}
\usepackage{color}
\usepackage{morefloats}
\usepackage{rotating}
\usepackage{srcltx}
\usepackage{slashed}
\usepackage{subfigure}
\usepackage{multirow}
\usepackage{verbatim}
\usepackage{hyperref}
\usepackage{tabularx}
\usepackage{adjustbox}

\usepackage{overpic}
\usepackage{makecell}

\begin{document}

\title{ Productions of $X(3872)$/$Z_c(3900)$ and $X_2(4013)$/$Z_c(4020)$  in   $Y(4220)$ and $Y(4360)$ decays    }

\author{Ming-Zhu Liu}
\affiliation{
Frontiers Science Center for Rare isotopes, Lanzhou University,
Lanzhou 730000, China}
\affiliation{ School of Nuclear Science and Technology, Lanzhou University, Lanzhou 730000, China}

\author{Xi-Zhe Ling}
\affiliation{Institute of High Energy Physics,
Chinese Academy of Sciences,
Beijing 100049, China}

\author{Li-Sheng Geng}~\email{lisheng.geng@buaa.edu.cn}
\affiliation{School of Physics, Beihang University, Beijing 102206, China}
\affiliation{Beijing Key Laboratory of Advanced Nuclear Materials and Physics, Beihang University, Beijing, 102206, China}
\affiliation{Peng Huanwu Collaborative Center for Research and Education, Beihang University, Beijing 100191, China}
\affiliation{Southern Center for Nuclear-Science Theory (SCNT), Institute of Modern Physics, Chinese Academy of Sciences, Huizhou 516000,  China}

\date{\today}
\begin{abstract}

The two excited vector charmonium states $Y(4220)$ and  $Y(4360)$ are difficult to be understood as pure  $c\bar{c}$ charmonium states. Since they are located close to the mass thresholds of $\bar{D}D_{1}$ and  $\bar{D}^*D_{1}$,  they can be viewed as $\bar{D}D_{1}$ and  $\bar{D}^*D_{1}$ molecules. Furthermore, recent studies indicated  that the exotic states  $X(3872)$/$Z_c(3900)$  and  $X_2(4013)$/$Z_c(4020)$ are the isoscalar/isovector $\bar{D}D^{*}$ and isoscalar/isovector $\bar{D}^*D^{*}$ molecules, respectively.  In this work, in the molecular picture, we employ the triangle diagram mechanism to study the productions of $ Z_{c}(3900) $ and $X(3872)$ in the pionic and radiative decays of $Y(4220)$, as well as their heavy-quark spin symmetry (HQSS) partners, i.e., the productions of $Z_{c}(4020) $ and $ X_2(4013)$ in the pionic and radiative decays of $Y(4360)$. Using the effective Lagrangian approach,  we obtain the ratios of the branching fractions $\mathcal{B}[Y(4360) \to Z_c(4020) \pi]/\mathcal{B}[Y(4220) \to Z_c(3900) \pi]=1.2{\pm 0.3}$ and $\mathcal{B}[Y(4360) \to X_2(4013) \gamma]/\mathcal{B}[Y(4220) \to X(3872) \gamma]=0.5{\pm 0.1}$, almost independent of model parameters, which indicate that the productions of $X_2(4013)$ and $Z_c(4020)$ in the radiative and pionic decays of $Y(4360)$  are likely to be measured in the future. The experimental studies of the predicted decay modes will help verify the molecular nature of $X(3872)$, $Z_c(3900)$, and $Y(4220)$. We hope the present work can stimulate experimental and further theoretical studies on these decay modes.

\end{abstract}


\maketitle

\section{Introduction}

In 2003, the Belle Collaboration observed a charmonium-like state $X(3872)$ in the $J/\psi \pi\pi $ mass distribution of the $B$ decay~\cite{Belle:2003nnu},  which was later confirmed by several other experiments~\cite{BaBar:2004iez,CDF:2003cab,D0:2004zmu,CMS:2013fpt,LHCb:2011zzp,BESIII:2013fnz}. In 2013, the LHCb Collaboration concluded that the quantum numbers of $X(3872)$ are $J^{PC}=1^{++}$~\cite{LHCb:2013kgk}. The mass of $X(3872)$ is  lower than the Goldfrey-Isgur model prediction by 90~MeV~\cite{Godfrey:1985xj}, and the ratios of $\mathcal{B}[X(3872)\to J/\psi  \pi \pi \pi]/\mathcal{B}[X(3872)\to J/\psi\pi\pi]\approx 1$~\cite{Belle:2005lfc,BaBar:2010wfc,BESIII:2019qvy} and $\mathcal{B}[B^+\to X(3872) K^+]/\mathcal{B}[B^0\to X(3872) K^0]\approx 0.5$~\cite{Belle:2011vlx} hint at large isospin breaking  if $X(3872)$ is regarded  as a conventional charmonium $\chi_{c1}(2P)$. Due to these anomalous properties, $X(3872)$ is widely viewed as an exotic state, triggering many theoretical and experimental studies. Identified  as a
$\bar{D}D^{\ast}$ molecule, the mass and isospin breaking decay modes of $X(3872)$ can be well described~\cite{Swanson:2003tb,Voloshin:2003nt,AlFiky:2005jd,Liu:2008fh,Sun:2011uh,Nieves:2012tt,Guo:2013sya,Karliner:2015ina,Liu:2019stu}, which imply that the $X(3872)$  contains a sizable $\bar{D}D^{\ast}$ molecular component. Considering the heavy-quark spin symmetry(HQSS), the contact potentials of $J^{PC}=1^{++}$  $\bar{D}D^{\ast}$ and  $J^{PC}=2^{++}$ $\bar{D}^*D^{\ast}$ systems  are the same~\cite{Liu:2020tqy}. Therefore, assuming $X(3872)$ as a $J^{PC}=1^{++}$  $\bar{D}D^{\ast}$ bound state, it is natural to expect  a $J^{PC}=2^{++}$ $\bar{D}^*D^{\ast}$ bound state (denoted by $X_2(4013)$), in agreement with Refs.~\cite{Nieves:2012tt,Guo:2013sya,Baru:2016iwj}. The BESIII Collaboration did not find evidence for $X_2(4013) $ by studying the process $e^+e^- \to \rho^0 X_{2}(4013)$ following the decay $X_2(4013) \to D\bar{D}$~\cite{BESIII:2019tdo}, and the  Belle Collaboration discovered  a similar state named as $X_2(4014)$ in the $\gamma\psi(2S)$ mass distribution of the  $\gamma\gamma\to \gamma\psi(2S)$ process with a low  statistical significance~\cite{Belle:2021nuv}. 

In 2013, the BESIII Collaboration and Belle Collaboration discovered a charged charmonium-like state $Z_{c}(3900)$ in the $J/\psi \pi^{\pm}$ mass distribution of $e^{+}e^{-}\to J/\psi \pi^{+}\pi^{-}$~\cite{BESIII:2013ris,Belle:2013yex}, which is above the $\bar{D}D^{\ast}$ mass threshold. As a result, the $Z_{c}(3900)$ can be explained as a $\bar{D}D^{\ast}$ resonant state. In terms of HQSS, the contact-range potentials of $J^{PC}=1^{+-}$ $\bar{D}D^{\ast}$ and $J^{PC}=1^{+-}$ $\bar{D}^{\ast}D^{\ast}$ are the same~\cite{Yang:2020nrt,Wu:2023rrp},  and therefore it is natural to expect the existence of a $\bar{D}^{\ast}D^{\ast}$ resonant state, which may be associated with the $Z_{c}(4020)$ state discovered by the BESIII Collaboration in the $\pi^{\pm}h_{c}$ mass distribution of $e^{+}e^{-}\to h_{c} \pi^{+}\pi^{-}$ ~\cite{BESIII:2013ouc}. In Ref.~\cite{Wang:2020dko}, Wang et al. concluded  that  $Z_{c}(3900)$ and $Z_{c}(4020)$ are $\bar{D}D^{\ast}$ and $\bar{D}^{\ast}D^{\ast}$ resonant states related by the HQSS.  This molecular doublet picture was reinforced by a series of studies after $Z_{cs}(3985)$~\cite{BESIII:2020qkh}, the SU(3) flavor partner of $Z_{c}(3900)$, was discovered~\cite{Yang:2020nrt,Meng:2020ihj,Baru:2021ddn,Yan:2021tcp,Du:2022jjv}.

The vector charmonium states can be produced directly in $e^+ e^-$ collisions. Among them,  $Y(4260)$ has been intensively studied, which was first discovered by the BaBar Collaboration in the $J/\psi \pi \pi$ mass distribution of the $e^+ e^- \to \gamma_{ISR} \pi^+ \pi^- J/\psi$ process~\cite{BaBar:2005hhc},  and then confirmed by the CLEO Collaboration~\cite{CLEO:2006ike} and Belle Collaboration~\cite{Belle:2007dxy}. Later, with larger data samples, the BESIII Collaboration found that the $Y(4260)$ split into two states $Y(4220)$ and $Y(4320)$~\cite{BESIII:2016bnd},  in agreement with the previous observations of $Y(4260)$ and $Y(4360)$, where $Y(4360)$ is discovered in the process of $e^+ e^- \to  \pi^+ \pi^- \psi(2S)$~\cite{BaBar:2006ait}.   Referring to the Review of Particle Physics(RPP)~\cite{ParticleDataGroup:2020ssz}, the masses and widths of $Y(4220)$ and $Y(4360)$ are $(4222, 48)$~MeV and $(4372, 115)$~MeV, respectively. 
From the measured  $R$ value of the BESIII Collaboration, the peaks at the energies of  4220 MeV and 4360 MeV are not pronounced~\cite{BES:2007zwq}, which indicate that these two states can not be easily interpreted as conventional charmonium states.   {According to a recent analysis of the charmonium spectrum in the unquenched quark model~\cite{Deng:2023mza},  $Y(4220)$ and $Y(4360)$ can not be assigned as excited   $S$-wave or $D$-wave conventional charmonium states. These properties favor  $Y(4220)$ and $Y(4360)$  as exotic states.    
}    
Since $Y(4220)$ and $Y(4360)$ are near the mass thresholds of $\bar{D}D_{1}$ and $\bar{D}^* D_1$, they can be viewed as  $\bar{D}D_{1}$ and $\bar{D}^* D_1$ molecules.  Recently, Ji et al. employed the meson exchange theory to assign $Y(4220)$, $Y(4360)$, and $Y(4415)$ as 
$\bar{D}D_{1}$, $\bar{D}^* D_1$, and $\bar{D}^* D_2$ molecules~\cite{Ji:2022blw},  confirmed later by a similar approach~\cite{Wang:2023ivd}. In addition to  the molecular interpretation for the $XYZ$ states, there exist other interpretations such as compact tetraquark states,  mixing states, and kinematic effects, see, e.g., Refs.~\cite{Chen:2016qju,Hosaka:2016ypm,Lebed:2016hpi,Oset:2016lyh,Esposito:2016noz,Dong:2017gaw,Guo:2017jvc,Olsen:2017bmm,Ali:2017jda,Karliner:2017qhf,Guo:2019twa,Brambilla:2019esw,Liu:2019zoy,Meng:2022ozq,Liu:2024uxn}.

It should be noted that the production of $X$ and $Z$ states are correlated with the excited $Y$ states.  In Ref.~\cite{BESIII:2013fnz}, the BESIII Collaboration found that $X(3872)$ can be produced through the radiative decay of   $Y(4260)$.  Moreover, the  $Z_{c}(3900)$  can be produced in the decay of $Y(4260)$~\cite{BESIII:2013ris,Belle:2013yex}. The experimental measurements have motivated theoretical studies on the production of the $X$ and $Z$ states in the $Y$ decay.   
In Refs.~\cite{Wang:2013cya,Chen:2013coa,Liu:2013vfa}, the authors described the line shape of $J/\psi \pi$ and $\pi\pi$ mass distribution and reproduced the $Z_{c}(3900)$  peak in the decay of $Y(4260) \to J/\psi \pi \pi$ with the assumption that $Y(4260)$ is strongly coupled to a pair of ground and excited charmed mesons.
In Ref.~\cite{Guo:2013zbw}, Guo et al. proposed to produce $X(3872)$ in the most promising channel $Y(4260)\to X(3872)\gamma$, where $Y(4260)$ and $X(3872)$ are proposed as $\bar{D}D_{1}$ and $\bar{D}D^*$ molecules. In the molecular picture, the decay width of  $Y(4260)\to X(3872)\gamma$ is estimated to be tens of keV through the effective Lagrangian approach~\cite{Dong:2014zka}.  Later,  Guo et al. investigated the production of  $X_2(4013)$ in the $Y(4260)$ radiative decay~\cite{Guo:2014ura}.  Chen et al. 
estimated the  partial  decay widths of  $Y(4390)$ assuming  $Y(4390)$ as a $\bar{D}^*D_{1}$ molecule, where the $Z_c(4020)$ is  produced in the $Y(4390)$ decay~\cite{Chen:2017abq}.

\begin{figure}[ttt]
\begin{center}
\begin{overpic}[scale=1.0]{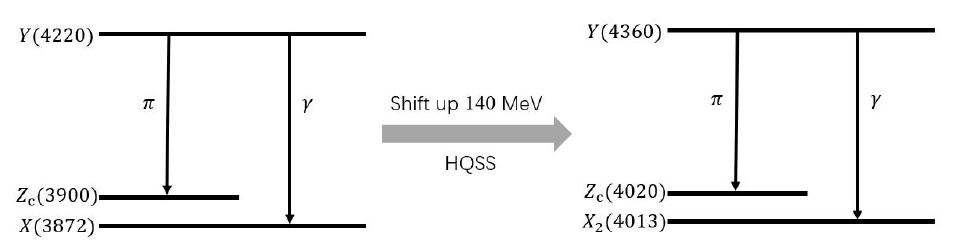}
\end{overpic}
\caption{ Decays of  $Y(4220) \to Z_{c}(3900)\pi / X(3872)\gamma $ and  decays   of their HQSS partners  $Y(4360) \to Z_{c}(4020)\pi / X_2(4013)\gamma $.  }
\label{hqss1}
\end{center}
\end{figure}

The productions of $X(3872)$ and $Z_c(3900)$ in the decays of $Y(4260)$ inspire us to study the relationship among the $XYZ$ states.  
As indicated in  Ref.~\cite{Zhu:2021vtd}, there is some relation among $X(3872)$,  $Z_c(3900)$, and  $Y(4220)$  as tetraquark states that can be seen on the triangle diagram, where  $X(3872)$ and  $Z_c(3900)$ are produced in the radiative and pionic decays of $Y(4220)$. Two other sets of  $XYZ$ states can be connected via the triangle diagram  with a similar relationship, i.e., $Y(4360)$,  $X(4012)$, and $Z_c(4020)$ as well as   $Y(4590)$, $X(4232)$, and $Z_{cs}(4260)$.  The former three masses can be obtained by shifting the masses of $X(3872)$,  $Z_c(3900)$, and  $Y(4220)$ up by $140$ MeV and the latter by $360$ MeV. We note that the so-obtained masses of the former three agree more with the corresponding experimental data. Therefore, in this work, we only focus on the $X(3872)$, $Z_c(3900)$, and $Y(4220)$ and $X_2(4013)$, $Z_c(4020)$, and $Y(4360)$,  assuming   $Y(4220)$ and  $Y(4360)$ as  $\bar{D}D_{1}$ and  $\bar{D}^* D_1$ bound states, $X(3872)$ and $X_2(4013)$ as $\bar{D}D^*$ and $\bar{D}^*D^*$ bound states, and  $Z_{c}(3900)$ and $Z_{c}(4020)$ as $\bar{D}D^*$ and $\bar{D}^*D^*$ resonant states. More specifically,  we explore their relationship via the radiative and pionic decays of the $Y$ states as shown in Fig.~\ref{hqss1}, where two sets of  $XYZ$ states are related to each other by HQSS.    

This work is organized as follows. We
briefly introduce the triangle diagram mechanism for the $Y(4220)/Y(4360)$ as the $\bar{D}D_{1}/\bar{D}^* D_1$ molecules  decaying into  $Z_{c}(3900)/Z_{c}(4020)$ as the  $\bar{D}D^*/\bar{D}^*D^*$ molecules and $\pi(\gamma)$ as well as $X(3872)/X_2(4013)$ together with $\pi$ and $\gamma$, and the effective Lagrangian approach in Sec.~II. The numerical results and discussions are given in Sec.~III, followed by a summary in the last section.

\section{Theoretical formalism}

\begin{figure}[!h]
\begin{center}
\begin{overpic}[scale=0.55]{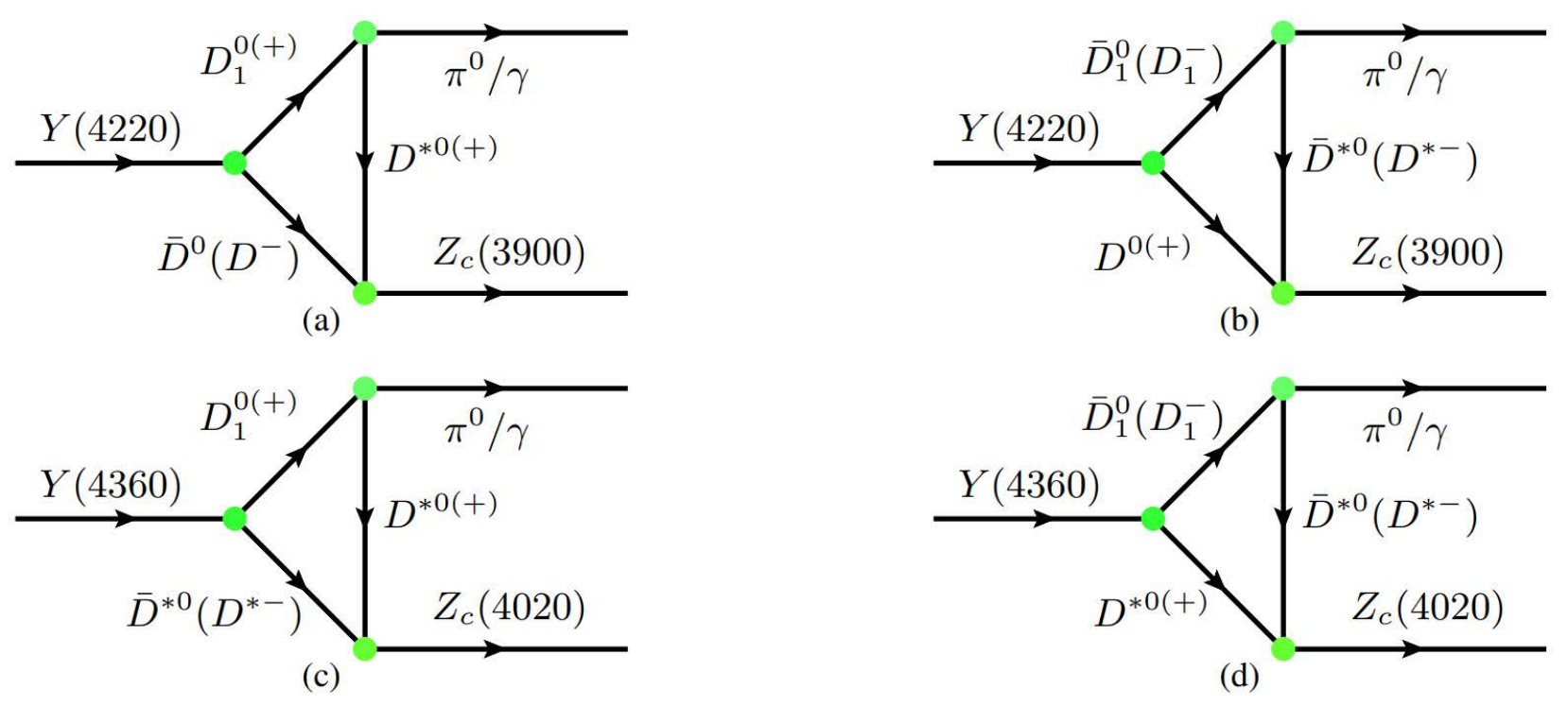}
\end{overpic}
\caption{ Triangle diagrams accounting for $ Y(4220) \to Z_c(3900)~ \pi^0 /\gamma$~(a,b)  and  $ Y(4360) \to Z_c(4020) ~\pi^0/ \gamma$~(c,d) in the molecular picture. }
\label{triangle1}
\end{center}
\end{figure}

\begin{figure}[!h]
\begin{center}
\begin{overpic}[scale=0.55]{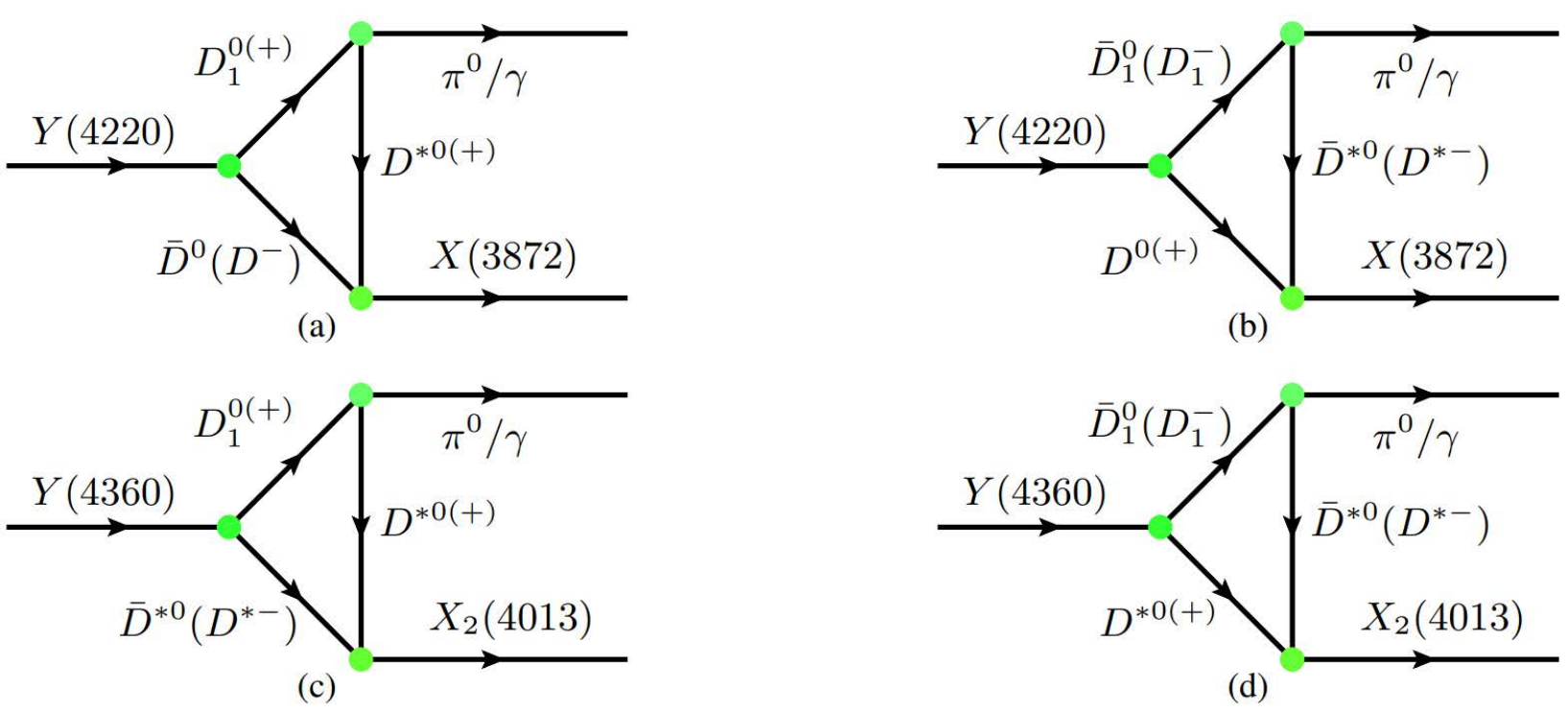}
\end{overpic}
\caption{   Triangle diagrams accounting for $ Y(4220) \to  X(3872)~ \pi^0 /\gamma$~(a,b)  and  $ Y(4360) \to X_2(4013) ~\pi^0/ \gamma$~(c,d) in the molecular picture.  }
\label{triangle2}
\end{center}
\end{figure}

First, we explain how to construct the triangle diagrams to study the radiative and pionic decays of $Y(4220)$ and $Y(4360)$ to $ Z_{c}(3900)/X(3872)$ and $ Z_{c}(4020)/X_2(4013)$. In the molecular picture, the triangle diagrams for the radiative and pionic decays of $Y(4220)$ and $Y(4360)$ are shown in Figs.~\ref{triangle1} and \ref{triangle2}, where the excited charmed meson $D_{1}$  first decays into the ground-state mesons $ {D}^{(\ast)}$ and $\pi(\gamma)$,  and then the $\bar{D}^{(\ast)}$ and ${D}^{(\ast)}$ interactions dynamically generate the molecules  $X(3872)$,  $X_2(4013)$, $Z_{c}(3900)$, and $Z_{c}(4020)$.  One can see that the initial and final molecules in Figs.~\ref{triangle1} (a-b) and Figs.~\ref{triangle1} (c-d) as well as in Figs.~\ref{triangle2} (a-b) and Figs.~\ref{triangle2} (c-d) are related to each other by the HQSS. Therefore,  using the decays of $Y(4220) \to X(3872) \gamma$ and  $Y(4220) \to Z_c(3900) \pi$ already measured experimentally as a benchmark, we can predict the decays of their HQSS partners, i.e.,  $Y(4360) \to X_2(4012) \gamma$ and  $Y(4360) \to Z_c(4020) \pi$.  

\subsection{ Effective Lagrangians }

In this work, we employ the effective Lagrangian approach to calculate the Feynman diagrams of Fig.~\ref{triangle1} and Fig.~\ref{triangle2}. The Lagrangians that describe each interaction vertex in the triangle diagrams can be classified into two categories.  
The first category involves the radiative and pionic decays of the excited charmed meson, i.e., $D_{1}\to D^{(\ast)}\gamma$ and $D_{1}\to D^{(\ast)}\pi$. The orbital angular momentum allowed for the decay $D_{1}\to D^{\ast} \pi$ is either $L=0$ or $L=2$.  Being the HQSS partner of $D_1$, $D_2$ can only decay into $D^*\pi$ via $D-$wave, and the decay of $D_{1}\to D^{\ast} \pi$ only proceeds via $D-$wave as well, which is the main reason why $D_{1}$ and $D_2$ are narrow.  

The Lagrangians describing the decays of $D_{1}\to D^{\ast} \pi$ and  $D_{1}\to D^{(\ast)} \gamma$  read~\cite{Ding:2008gr,Li:2013yla,Liu:2013vfa,Dong:2014zka} 
\begin{eqnarray}
\mathcal{L}_{D_{1} D^* \pi}&=&  g_{D_{1} D^* \pi}(3D_{1}^{\mu}\partial_{\mu}\partial_{\nu}\pi D^{\ast\nu}- D_{1}^{\mu}\partial^{\nu}\partial_{\nu}\pi D_{\mu}), 
\\ \nonumber
\mathcal{L}_{D_{1} D^* \gamma}&=& e g_{D_{1} D^* \gamma} \varepsilon^{\mu\nu\alpha\beta}\partial_{\mu}A_{\nu}D_{1\alpha} D^{\ast}_{\beta},
\end{eqnarray}
where the couplings $g_{D_{1} D^* \pi}$ and $g_{D_{1} D^* \gamma}$ can be determined by saturating their partial widths~\cite{Cleven:2013mka}.  
We assume that the decay mode $D_{1}\to D^* \pi$ saturates the width of $D_{1}$. Therefore, one can determine the coupling $g_{D_{1}D^* \pi}=12.67$~GeV$^{-1}$, consistent with Refs.~\cite{Li:2013yla,Wang:2021yld}.   On the other hand, $D_{1}(2420)$, viewed as a conventional  $^1P_{1}$ state, decays into $D\gamma$ via the electromagnetic transition.  As for the other state with the same angular momentum, the $D_{1}(2430)$, viewed as a conventional  $^3P_{1}$ state, only decays into $D^*\gamma$. However, the physical $D_{1}(2420)$ and $D_{1}(2430)$ states are mixtures of $^3P_{1}$ and $^1P_{1}$ components with a mixing angle in the range of $-25.7^{\circ} \sim -54.7^{\circ}$ ~\cite{Godfrey:2015dva,Li:2022vby}. In this work, we take the average value of $D_{1}(2420)$ radiative decay width obtained in Refs.~\cite{Godfrey:2015dva,Li:2022vby} to fix the couplings.   
Then, with the widths of $\Gamma_{D_{1}^0\to D^{\ast0} \gamma}=187$ keV and $\Gamma_{D_{1}^+\to D^{\ast+} \gamma}=19$ keV, we obtain the couplings: $g_{D_{1}^0  D^{\ast0} \gamma}=1.82$ and $g_{D_{1}^+  D^{\ast+} \gamma}=0.58$, with $e=0.303$.   
 
 The second category involves the interactions between the hadronic molecules and their constituents. 
The molecules' interactions with the corresponding constituents are described by the following Lagrangians~\cite{Li:2013yla,Dong:2013iqa,Chen:2017abq} 
\begin{eqnarray}
\mathcal{L}_{Y(4220)D_{1}\bar{D}}&=&  g_{Y(4220)D_{1}\bar{D}} Y^{\nu}(4220)D_{1\mu}\bar{D},    \\ \nonumber
\mathcal{L}_{Y(4360)D_{1}\bar{D}^*}&=&  g_{Y(4360)D_{1}\bar{D}^*} \epsilon^{\mu\nu\alpha\beta}\partial_{\mu}Y_{\nu}(4360)D_{1\alpha}\bar{D}^{*}_{\beta},    \\ \nonumber
\mathcal{L}_{Z_{c}(3900)D\bar{D}^*}&=&  g_{Z_{c}(3900)D\bar{D}^*} Z_{c}^{\mu}(3900)D\bar{D}^*_{\mu},  \\ \nonumber
\mathcal{L}_{Z_{c}(4020)D^*\bar{D}^*}&=& i g_{Z_{c}(4020)D^*\bar{D}^*}\varepsilon_{\mu\nu\alpha\beta} \partial^{\mu}Z_{c}^{\nu}(4020)D^{\ast\alpha}\bar{D}^{\ast\beta},
\\ \nonumber
\mathcal{L}_{X(3872)D\bar{D}^*}&=&  g_{X(3872)D\bar{D}^*} X^{\mu}(3872)D\bar{D}^*_{\mu}, 
\\ \nonumber
\mathcal{L}_{X_2(4013)D^*\bar{D}^*}&=&  g_{X_2(4013)D^*\bar{D}^*} X_2^{\mu\nu}(4013)D^*_{\mu}\bar{D}^*_{\nu}, 
\end{eqnarray}
where  $g$ with the specific subscript denotes the molecule's couplings to their constituents, estimated in the contact-range approach.   

\subsection{ Contact-range effective field theory(EFT) }

These couplings are estimated from the residues of poles obtained by solving the Lippmann-Schwinger equation~\cite{Liu:2019tjn}
\begin{eqnarray}
T=(1-VG)^{-1}V,
\end{eqnarray}
where $V$ is the hadron-hadron potential determined by the contact EFT approach described below, and $G$ is the two-body propagator. 
In evaluating the loop function $G$, we introduce a regulator of Gaussian form $e^{-2q^{2}/\Lambda^2}$ in the integral as
\begin{eqnarray}
G(s)=\int \frac{d^{3}q}{(2\pi)^{3}} \frac{e^{-2q^{2}/\Lambda^2}}{{\sqrt{s}}-m_{1}-m_{2}-q^{2}/(2\mu_{12})+i \varepsilon}
\label{loopfunction},
\end{eqnarray}
where $\sqrt{s}$ is  the total energy in the center-of-mass frame of $m_{1}$ and $m_{2}$, $\mu_{12}=\frac{m_{1}m_{2}}{m_{1}+m_{2}}$ is the reduced mass, and $\Lambda$ is the momentum cutoff. Following our previous works~\cite{Liu:2019tjn,Xie:2022hhv}, we take $\Lambda=1$~GeV in the present work. As the poles are dynamically generated in the unphysical sheet, the loop function of Eq.~(\ref{loopfunction}) becomes
\begin{eqnarray}
G^{II}(s,m_1,m_2)=G^{I}(s,m_1,m_2)+i\mu_{12} \frac{p  }{2\pi}{  e^{-2 p^2/\Lambda^2}},
\end{eqnarray}
where the c.m. momentum $p$  is  
\begin{eqnarray}
p=\sqrt{2\mu_{12}\left(\sqrt{s}-m_1-m_2\right)}.
\end{eqnarray}

In the heavy-quark limit, the $\bar{D}^{(*)}D^{(*)}$ contact potentials are characterized by two parameters $C_a$ and $C_b$ ( see Ref.~\cite{Liu:2020tqy} for details), and the $\bar{D}^{(*)}D_1$ contact potentials are characterised by four parameters $C_a$, $C_b$, $D_a$, and $D_b$ (see Ref.~\cite{Peng:2022nrj} for details). { In the language of EFTs, the expansion parameter for the $\bar{D}D_1$ system can be parameterized as the ratio of the binding momentum $\gamma=\sqrt{2\mu B}$ (where $\mu$ is the reduced mass of the $\bar{D}D_1$ system and $B$ is the corresponding binding energy) to the $\rho$ meson mass, i.e., $\gamma/m_{\rho}\sim 0.5$, which still lies within the validity range of EFTs but may suffer from slow convergence. } Here, we only present the potentials to be used in the present work 
\begin{eqnarray}
V(J^{PC}=1^{++}~\bar{D}^*D)&=&C_a+C_b,  \\ \nonumber 
V(J^{PC}=2^{++}~\bar{D}^*D^*)&=&C_a+C_b, \\ \nonumber 
V(J^{PC}=1^{+-}~\bar{D}^*D)&=&C_a-C_b, 
\\ \nonumber 
V(J^{PC}=1^{+-}~\bar{D}^*D^*)&=&C_a-C_b, 
\\ \nonumber 
V(J^{PC}=1^{--}~\bar{D}D_1)&=&C_a-\frac{2}{3}D_b, 
\\ \nonumber 
V(J^{PC}=1^{--}~\bar{D}^*D_1)&=&C_a-\frac{5}{4}C_b-\frac{1}{6}D_b-\frac{5}{4}D_c. 
\end{eqnarray}

With the  potentials above,  we  search for poles in the vicinity of  $\bar{D}^{(*)}D^{(*)}$ and $\bar{D}^{(*)}D_1$ mass thresholds and  then determine the 
 couplings between the molecular states and their constituents from the residues of the corresponding poles, 
\begin{eqnarray}
g_{i}g_{j}=\lim_{{\sqrt{s}}\to {\sqrt{s_0}}}\left({\sqrt{s}}-{\sqrt{s_0}}\right)T_{ij}(\sqrt{s}),
\end{eqnarray}
where $g_{i}$ denotes the coupling of channel $i$ to the  dynamically generated state and ${\sqrt{s_0}}$ is the pole position. 

\subsection{ Decay amplitudes  }
 
Utilizing all the relevant Lagrangians, the amplitudes of the decays $Y(4220)[k_0,\varepsilon(k_0)]\to  Z_{c}(3900)[p_2,\varepsilon(p_2)]\pi(p_1)$ and $Y(4360)[k_0,\varepsilon(k_0)]\to  Z_{c}(4020)[p_2,\varepsilon(p_2)]\pi(p_1)$ in Fig.~\ref{triangle1}
 can be expressed as 
 \begin{eqnarray}
 i\mathcal{M}_{1a,1b}^{\pi}&=&g_{Y(4220)D_1\bar{D}}g_{D_1 D^* \pi} g_{D^* \bar{D} Z_{c}(3900)}\int \frac{d^{4}q}{(2\pi)^4}  \frac{1}{q_1^2-m_{D_1}^2}\frac{1}{q_{2}^{2}-m_{\bar{D}}^{2}}\frac{1}{q^{2}-m_{D^*}^{2}}\varepsilon^{\mu}(k_0) \\   \nonumber
 &&(-g^{\mu\nu}+\frac{q_{1}^{\mu}q_{1}^{\nu}}{m_{D_1}^2})(3p_{1}^{\nu}p_{1}^{\alpha}-g^{\nu\alpha}p_{1}^2)(-g^{\alpha\beta}+\frac{q^{\alpha}q^{\beta}}{m_{D^*}^2}) \varepsilon^{\beta}(p_{2})  F(q^2),  \\  \nonumber
 i\mathcal{M}_{1c,1d}^{\pi}&=&g_{Y(4360)D_1^*\bar{D}}g_{D_1 D^* \pi} g_{D^* \bar{D}^* Z_{c}(4020)}\int \frac{d^{4}q}{(2\pi)^4}  \frac{1}{q_1^2-m_{D_1}^2}\frac{1}{q_{2}^{2}-m_{\bar{D}^*}^{2}}\frac{1}{q^{2}-m_{D^*}^{2}}\varepsilon^{\mu\nu\alpha\beta} k_{0\mu}\varepsilon_{\nu}(k_0)\\ \nonumber 
 &&(-g^{\alpha\theta}+\frac{q_{1}^{\alpha}q_{1}^{\theta}}{m_{D_1}^2})  (3p_{1}^{\theta}p_{1}^{\tau}-g^{\theta\tau}p_{1}^2)(-g^{\tau\rho}+\frac{q^{\tau}q^{\rho}}{m_{D^*}^2})(-g^{\beta\phi}+\frac{q_2^{\beta}q_2^{\phi}}{m_{D^*}^2}) \varepsilon^{\omega\eta\phi\rho}p_{2\omega}\varepsilon_{\eta}(p_{2})  F(q^2),   
 \label{pi}
 \end{eqnarray}
 where  
$q_{1}$, $q_{2}$, and $q$  denote the momenta of  $D_{1}$, $\bar{D}^{(\ast)}$, and $ {D}^{\ast}$ in the triangle diagrams, and  $\varepsilon^{\mu}$ and $\varepsilon^{\mu\nu}$ represent the polarization vectors for the states of spin $S=1$ and  spin  $S=2$, respectively.   The tensor polarization vector  satisfies the following relationships
\begin{eqnarray}
\varepsilon^{\mu\nu}_{(\lambda)}(q,m) q_{\mu} &=& 0       \\ \nonumber
\varepsilon^{\mu\nu}_{(\lambda)}(q,m) g_{\mu\nu} &=& 0       \\ \nonumber
\sum_{\lambda=0,\pm 1,\pm2}\varepsilon^{\mu\nu}_{(\lambda)}(q,m)\varepsilon^{\mu^{\prime}\nu^{\prime}}_{(\lambda)}(q,m) &=&  \frac{1}{2}(\bar{g}_{\mu\mu^{\prime}}\bar{g}_{\nu\nu^{\prime}}+ \bar{g}_{\mu\nu^{\prime}}\bar{g}_{\nu\mu^{\prime}})-\frac{1}{3}\bar{g}_{\mu\nu}\bar{g}g_{\mu^{\prime}\nu^{\prime}},     
\end{eqnarray}
with $ \bar{g}^{\mu\nu}= -g^{\mu\nu}+\frac{q^{\mu}q^{\nu}}{q^2}  $.

Similarly, we express the amplitudes  of   the decays $Y(4220)[k_0, \varepsilon(k_0)]\to  X(3872)[p_1, \varepsilon(p_1)] \,\gamma[p_2, \varepsilon(p_2)]$ and $Y(4360)[k_0, \varepsilon(k_0)]\to  X(4013)[p_1, \varepsilon(p_1)] \,\gamma[p_2, \varepsilon(p_2)]$ of Fig.~\ref{triangle2} 
 \begin{eqnarray}
 \label{gamma}   
i\mathcal{M}_{2a,2b}^{\gamma}&=&g_{Y(4220)D_1\bar{D}}g_{D_1 D^* \gamma} g_{D^* \bar{D} X(3872)}\int \frac{d^{4}q}{(2\pi)^4}  \frac{1}{q_1^2-m_{D_1}^2}\frac{1}{q_{2}^{2}-m_{\bar{D}}^{2}}\frac{1}{q^{2}-m_{D^*}^{2}}  \varepsilon_{\mu}(k_0)\\    \nonumber
 &&(-g^{\mu\nu}+\frac{q_{1}^{\mu}q_{1}^{\nu}}{m_{D_1}^2}) \varepsilon^{\alpha\beta\nu\theta}p_{1\alpha}\varepsilon_{\beta}(p_{1}) (-g^{\theta\phi}+\frac{q^{\theta}q^{\phi}}{m_{D^*}^2}) \varepsilon_{\phi}(p_{2})  F(q^2),   \\  \nonumber
 i\mathcal{M}_{2c,2d}^{\gamma}&=&g_{Y(4360)D_1\bar{D}^*}g_{D_1 D^* \gamma} g_{D^* \bar{D}^* X(4013)}\int \frac{d^{4}q}{(2\pi)^4}  \frac{1}{q_1^2-m_{D_1}^2}\frac{1}{q_{2}^{2}-m_{\bar{D}^*}^{2}}\frac{1}{q^{2}-m_{D^*}^{2}} \varepsilon^{\mu\nu\alpha\beta}  \\  \nonumber
 && k_{0\mu}\varepsilon_{\nu}(k_0)(-g^{\alpha\phi}+\frac{q_{1}^{\alpha}q_{1}^{\phi}}{m_{D_1}^2}) \varepsilon^{\omega\eta\phi\rho}p_{1\omega}\varepsilon_{\eta}(p_{1}) (-g^{\rho c}+\frac{q^{\rho}q^{c}}{m_{D^*}^2})   (-g^{\beta d}+\frac{q_2^{\beta}q_2^{d}}{m_{D^*}^2})\varepsilon_{ c d}    F(q^2).  
 \end{eqnarray}

   In addition, to eliminate the ultraviolet divergence of the above amplitudes, we  supplement the relevant vertices of ${D}^*$ exchange with the following  monopolar form factor $F(q^2)$
 \begin{eqnarray}
 F(q^2)=(\frac{{\Lambda_m}^2-m^2}{{\Lambda_m}^2-q^2})^2 ,
 \end{eqnarray}
 which reflects the internal structure of hadrons, similar to the OBE model~\cite{Liu:2018bkx}.  { $m$ represents the mass of the exchanged particles in the triangle diagrams.   }   An  additional parameter ${\Lambda_m}$ is parameterised  as a dimensionless  parameter $\alpha$, i.e.,   ${\Lambda_m}=m+\alpha \Lambda_{QCD}$,  where $\Lambda_{QCD}\sim 200-300$ MeV, and $\alpha$ is around $1$.  In this work, we vary $\alpha$ from $0.5$ to $1.5$ to study the result's dependence on the cutoff.

For the radiative decays, gauge symmetry must be satisfied.        
If the quantum number of  the initial state is $1^{-}$ and that of the final state is  $1^{+}$, the covariant   amplitude of the radiative  decay for the triangle diagram  is written as 
 \begin{eqnarray}
\mathcal{M} = \varepsilon_{\mu}(k_0)\varepsilon_{\beta}(p_1)\varepsilon_{\phi}(p_2)M_{loop}^{\mu\beta\phi},
 \end{eqnarray}
where $M_{loop}^{\mu\beta\phi}$   contains the following independent terms
 \begin{eqnarray}
M_{loop}^{\mu\beta\phi}=\varepsilon^{p_2p_1\mu\phi}p_1^{\beta} g_1^{Tri}(q^2)
                       +  \varepsilon^{p_2p_1\beta\phi}p_2^{\mu}g_2^{Tri}(q^2),
                       \label{4220gamma}
 \end{eqnarray}
with $g^{Tri}(q^2)$ representing the loop integrals~\cite{Dubnicka:2011mm,Dong:2014zka,Ganbold:2021nvj}.  
 
If the quantum number of  the initial state is $1^{-}$ and that of the final state is  $2^{+}$, the corresponding amplitude is  expressed  as
 \begin{eqnarray}
\mathcal{M} = \varepsilon_{\nu}(k_0)\varepsilon_{\eta}(p_1)\varepsilon_{cd}(p_2)M_{loop}^{\nu\eta c d },
 \end{eqnarray}
where  $M_{loop}^{\nu\eta c d }$ is written as the following independent  terms~\cite{Ganbold:2021nvj} 
 \begin{eqnarray}
 \label{4360gamma}
M_{loop}^{\nu\eta c d }&=& g_1^{Tri}(q^2)\{ g^{c\nu}[g^{\eta d}(p_1\cdot p_2)-p_1^d p_2^\eta ] +g^{d\nu}[g^{\eta c}(p_1\cdot p_2)-p_1^c p_2^\eta ] \}   \\ \nonumber &+&  g_2^{Tri}(q^2) \{ g^{\eta\nu}[p_1^c p_2^d + p_1^d p_2^c ] -g^{c\eta} p_1^{\nu} p_2^d -g^{d\eta} p_1^{\nu} p_2^{c}   \}.    
 \end{eqnarray}

 With the amplitudes of pionic and radiative decays of the $Y$ states to the $X$ and $Z$ states determined, one can obtain 
 the corresponding partial decay width as
 \begin{eqnarray}
\Gamma=\frac{1}{2J+1}\frac{1}{8\pi}\frac{|\vec{p}|}{m_{Y}^2}\bar{|\mathcal{M}|}^{2},
\end{eqnarray}
where $J$ is the total angular momentum of the initial $Y$ state, the overline indicates the sum over the polarization vectors of the final states, and $|\vec{p}|$ is the momentum of either final state in the rest frame of the $Y$ state.

\section{Numerical Results and Discussions}
\label{sec:Results}

\begin{table}[!h]
\caption{Masses and quantum numbers of the mesons relevant to the present work.~\cite{ParticleDataGroup:2020ssz}. \label{mass}}
\begin{tabular}{ccc|ccc}
  \hline\hline
   Meson & $I (J^P)$ & M (MeV) &    Meson & $I (J^P)$ & M (MeV)   \\
  \hline  $\pi^{0}$ & $1(0^-)$ & $134.98$  &    $\pi^{\pm}$ & $1(0^-)$ & $139.57$ \\
  $D^{0}$ & $\frac{1}{2}(0^-)$ & $1864.84$  &    $D^{\pm}$ & $\frac{1}{2}(0^-)$ & $1869.66$ \\
  $D^{\ast0}$ & $\frac{1}{2}(1^-)$ & $2006.85$ &  $D^{\ast\pm}$ & $\frac{1}{2}(1^-)$ & $2010.26$
  \\  $D_{1}^{0}$ & $\frac{1}{2}(1^+)$ & $2422.06$ &  $D_{1}^{\pm}$ & $\frac{1}{2}(1^+)$ & $2426.06$
  \\
     $Y(4220)$ & $0(1^{--})$ & $4222.7$ &  $Y(4360)$ & $0(1^{--})$ & $4372$
  \\
   $Z_{c}(3900)$ & $1(1^{+-})$ & $3887.1$ &  $Z_{c}(4020)$ & $1(1^{+-})$ & $4024.1$
  \\
     $X(3872)$ & $0(1^{++})$ & $3871.65$ &  $X(4012)$ & $0(2^{++})$ & $4013$
  \\
 \hline \hline
\end{tabular}
\label{tab:masses}
\end{table}

In Table~\ref{tab:masses}, we tabulate relevant particles' masses and quantum numbers.
First, we determine the couplings between the molecules and their constituents in the contact-range EFT approach. Following Refs.~\cite{Pan:2023hrk,Wu:2023rrp}, we determine   
 the unknown parameters of the contact-range potential by fitting to the experimental data, and then the molecules' coupling to their constituents are derived from the residues of the poles.  For the  $\bar{D}^{(\ast)}{D}^{(\ast)}$ system,  
  we can see  that the $J^{PC}=1^{++}$  $\bar{D}D^{\ast}$ potential is the same as the $J^{PC}=2^{++}$  $\bar{D}^{\ast}D^{\ast}$ potential. 
 Identifying the $X(3872)$ as a $J^{PC}=1^{++}$  $\bar{D}D^{\ast}$ bound state,  we determine  the value of  $C_{a}+C_{b}=-20.25$ GeV$^{-2}$, and then obtain the coupling  $g_{X \bar{D} D^{\ast}}=9.16$ GeV. Considering the HQSS,  we find a $\bar{D}^*D^*$  bound state with a mass of $4013$ MeV  denoted by $X_{2}(4013)$,  and then  its coupling is determined as   
$g_{X_{2} \bar{D}^* D^{\ast}}=10.87$ GeV. Using the Weinberg compositeness theorem~\cite{Wang:2013cya}\footnote{ For a bound state $Y$ below the mass threshold $m_1+m_2$,  the coupling is generally written as $g=\sqrt{16\pi(m_1+m_2)^2\sqrt{\frac{2(m_1+m_2-m_{Y})}{m_1m_2/(m_1+m_2)}} }$ in Weinberg's approach.  }, one can calculate these couplings as $g_{X \bar{D}^* D}=8.29$ GeV and $g_{X_{2} \bar{D}^* D^{\ast}}=9.00$ GeV, consistent with our estimations.

For the $\bar{D}^{(\ast)}D_{1}$ system,      
assuming   $Y(4360)$ and $Y(4220)$ 
as  $J^{PC}=1^{--}$  $\bar{D}^{\ast}D_{1}$ and $J^{PC}=1^{--}$  $\bar{D}D_{1}$ bound states,  
   we determine the $D_{1}\bar{D}^*$ and $D_{1}\bar{D}$  contact  potentials as  $C_a-\frac{5}{4}C_b-\frac{1}{6}D_b-\frac{5}{4}D_c=-32.27$ GeV$^{-2}$ and $C_a-\frac{2}{3}D_b=-34.65$ GeV$^{-2}$, which to some extent reflect the HQSS between the $D_{1}\bar{D}^*$ and $D_{1}\bar{D}$ systems. 
   With the obtained contact-range potentials, we determine the couplings of $g_{Y(4360)D_{1}\bar{D}^*}=6.97$  and $g_{Y(4220)D_{1}\bar{D}}=31.32$ GeV.    In terms of the Weinberg compositeness theorem, one can obtain the couplings of $g_{Y(4360)D_{1}\bar{D}^*}=4.11$ and   $g_{Y(4220)D_{1}\bar{D}}=18.29$ GeV~\cite{Li:2013yla,Chen:2017abq}, which are smaller than those of our estimations as shown in Table~\ref{couplingparticle}.  We note that the Weinberg compositeness theorem is only applicable to weakly bound states, while the binding energies of the $\bar{D}^{(\ast)}D_{1}$ molecules are about 70 MeV.

{  Since the mass splitting between the $\bar{D}^*D_1$ and $\bar{D}^*D_2$ thresholds is only $37$~MeV, we estimate the coupled-channel effect in the following.     In the heavy quark limit, the coupled-channel $\bar{D}^*D_1-\bar{D}^*D_2$  contact-range potentials are parameterised as   
\begin{equation}
    V_{\bar{D}^*D_1-\bar{D}^*D_2}^{J^{PC}=1^{--}}=\begin{pmatrix} 
      C_a-\frac{5}{4}C_b-\frac{1}{6}D_b-\frac{5}{4}D_c&  \frac{\sqrt{5}}{4}C_b-\frac{\sqrt{5}}{6}D_b+\frac{3\sqrt{5}}{4}D_c\\ \frac{\sqrt{5}}{4}C_b-\frac{\sqrt{5}}{6}D_b+\frac{3\sqrt{5}}{4}D_c& C_a-\frac{9}{4}C_b-\frac{1}{6}D_b+\frac{3}{4}D_c\end{pmatrix}. 
      \label{potential1}
\end{equation}          

{Since only two experimental
observables exist, i.e., the binding energies of the $Y(4220)$ and $Y(4360)$,
we simplify the potential in Eq.(\ref{potential1}) assuming that the $C_b$ and $D_c$ terms
responsible for the spin-spin interaction are
negligible} \footnote{   Based on the assumption that the potentials of  $J^{PC}=1^{--}\bar{D}^{\ast}D_1 $ and $J^{PC}=1^{--}\bar{D}^{\ast}D_2^* $ are the same in the heavy quark limit,   the parameters $C_{b}$  and $D_{c}$  in Eq.~(18) are perturbative.  } 
\begin{equation}
    V_{\bar{D}^*D_1-\bar{D}^*D_2}^{J^{PC}=1^{--}}=\begin{pmatrix} 
      C_a-\frac{1}{6}D_b& -\frac{\sqrt{5}}{6}D_b \\  -\frac{\sqrt{5}}{6}D_b & C_a -\frac{1}{6}D_b\end{pmatrix}, 
\end{equation}  
where the values of $C_a=-31.48$~GeV$^{-2}$ and $D_b=4.76$~GeV$^{-2}$ are determined by reproducing the masses of $Y(4220)$ and $Y(4360)$. For the single-channel case, we obtain two bound states below the $\bar{D}^*D_1$ and  $\bar{D}^*D_2$ mass thresholds with the masses being $4372$~MeV and $4408$~MeV, respectively, which are both below the $\bar{D}^*D_1$  mass threshold. Considering the coupled-channel effects, these two poles shift to $4370$~MeV and  $4410$~MeV. The $Y(4360)$ coupling to $\bar{D}^*D_1$ decreases from  $6.97$ to $6.76$, resulting in a $3\%$ uncertainty. This exercise shows that the $\bar{D}^*D_1-\bar{D}^*D_2$ coupled-channel effects have a minor impact on the pole position and coupling of $Y(4360)$.      
}

The leading-order contact potentials can only generate bound or virtual poles~\cite{Dong:2020hxe}. To generate resonant states, one has to add higher-order momentum-dependent contact potentials~\cite{Zhai:2022ied}.
In the heavy-quark limit,  the contact potentials of  $J^{PC}=1^{+-}$  $\bar{D}D^{\ast}$ and  $J^{PC}=1^{+-}$  $\bar{D}^{\ast}D^{\ast}$ are expressed as $C_{LO}+C_{NLO}~q^2$, where $C_{LO}=C_a-C_b$ and 
$q$ is the c.m. momentum. By reproducing the mass and width of $Z_{c}(3900)$, one can obtain the value of  $C_{LO}=-7.7$ GeV$^{-2}$ and  $C_{NLO}=-211$ GeV$^{-4}$, and then determine the coupling 
$g_{Z_{c}(3900)\bar{D}D^*}=7.10$~GeV. With the obtained values of $C_{LO}$ and  $C_{NLO}$, we predict the mass and width of the $\bar{D}^*D^*$ molecule, i.e., (4028,13) MeV, consistent with the RPP, and then determine the coupling $g_{Z_{c}(4020)\bar{D}^*D^*}=1.77$.

\begin{table}[ttt]
\centering
\caption{ Values of the $XYZ$ states as the hadronic molecules  couplings to their constituents obtained by contact-range EFT approach and  Weinberg’s compositeness theorem. \label{couplingparticle}
}
\begin{tabular}{c| c c c c c c c cccc}
\hline\hline
    &~~ $X(3872) $   &~~ $X_2(4013)$  &~~$Y(4220) $  &~~$Y(4360) $   &~~  $Z_c(3900) $  &~~  $Z_c(4020)$  
         \\ \hline
    contact-range EFT    &~~ $9.16$~GeV   &~~ $10.87$~GeV  &~~ $31.32$~GeV  &~~ $6.97$ &~~ $7.10$~GeV &~~ $1.77$ \\  \hline
    Weinberg’s compositeness    &~~ $8.29$~GeV   &~~ $9.00$~GeV  &~~ $18.29$~GeV  &~~ $4.11$ \\
\hline\hline
\end{tabular}
\end{table}

In Table~\ref{couplingparticle}, we collect the values of the $XYZ$ molecular candidates couplings to their constituents, which are determined in the isospin basis.  However,  the couplings in the particle basis shown in Fig.~\ref{triangle1} and Fig.~\ref{triangle2} are necessary for our study.            
Working in the isospin limit, we present the ratios of the couplings in the particle basis to those in the isospin basis in Table~\ref{couplingsr1}, which help us determine the values of the molecules couplings to their constituents in the particle basis. Moreover, the signs of ratios given  in Table~\ref{couplingsr1}
are consistent with the signs in the triangle diagrams, which determine the signs between the triangle diagrams.     
In the isospin limit, the amplitudes for the decays  $Y(4220)\to X(3872)\pi$ and  $Y(4220)\to Z_c(3900)\gamma$ as well as their HQSS partners  $Y(4360)\to X_2(4013)\pi$ and  $Y(4360)\to Z_c(4020)\gamma$ illustrated by the triangle diagrams are zero, and therefore we focus on the decays of $Y(4220)\to X(3872)\gamma$ and  $Y(4220)\to Z_c(3900)\pi$ as well as the decays of their  HQSS partners $Y(4360)\to X_2(4013)\gamma$ and  $Y(4360)\to Z_c(4020)\pi$ in this work.

\begin{table}[ttt]
    \centering
    \caption{Ratios of the couplings in the particle basis to the coupling in the isospin basis.}
    \begin{tabular}{c|cccc}
    \hline\hline
     Molecules    &~~~~${D}^{\ast+}D^-$  &~~~~${D}^{+}D^{\ast-}$ &~~~~${D}^{\ast0}\bar{D}^{0}$   &~~~~${D}^{0} \bar{D}^{\ast0}$ 
         \\ \hline
        $X(3872) $   &~~~~ $1/2$   &~~$-1/2$ &~~~~ $1/2$  &~~$-1/2$ \\
        $Z_c(3900)$   &~~~~ $1/2$    &~~~~ $1/2$  &~~$-1/2$  &~~$-1/2$   \\ \hline    
     Molecules    &~~~~${D}^{+}_{1}D^{(\ast)-}$  &~~~~${D}^{(\ast)+}D_{1}^{-}$ &~~~~${D}^{0}_{1}\bar{D}^{(\ast)0}$   &~~~~${D}^{(\ast)0} \bar{D}^{0}_{1}$ 
         \\ \hline
        $Y(4220) $   &~~ $-1/2$   &~~~~$1/2$ &~~ $-1/2$  &~~~~$1/2$ \\
        $Y(4360)$    &~~ $-1/2$   &~~~~$1/2$ &~~ $-1/2$  &~~~~$1/2$         
         \\ \hline  Molecules     &~~~~${D}^{\ast+}D^{\ast-}$ &~~~~${D}^{\ast0}\bar{D}^{\ast0}$ & &
         \\ \hline
        $X_2(4013) $   &~~~~ $1/\sqrt{2}$  &~~~~ $1/\sqrt{2}$  & &  \\
        $Z_c(4020)$  &~~~~ $1/\sqrt{2}$  &~~ $-1/\sqrt{2}$   &  & 
\\
    \hline \hline
    \end{tabular}
   \label{couplingsr1}
\end{table}

In our theoretical framework, an unknown dimensionless parameter $\alpha$ exists in the form factor. The value of $\alpha$ is assumed to be around 1, and therefore, $\alpha$ is varied from 0.5 to 1.5 to reflect the uncertainties induced by the form factors.  Moreover, the uncertainties of the couplings in the three vertices of the triangle diagrams lead to uncertainties in the final results. Following Ref.~\cite{Wu:2023rrp}, the molecule's couplings to their constituents lead to a $10\%$ uncertainty as the cutoff $\Lambda$ varies from $1$ GeV to $2$ GeV. The HQSS is adopted to estimate the coupling $g_{D_1D^*\pi}$, resulting in a  $15\%$ uncertainty~\cite{Pan:2019skd}.  The coupling $g_{D_1D^*\gamma}$ likely leads to about a $30\%$ uncertainty due to averaging the widths of  Ref.~\cite{Godfrey:2015dva} and Ref.~\cite{Li:2022vby}.  Finally,  we obtain the uncertainties for the partial decay widths 
originating from the uncertainties of these couplings via a
Monte Carlo sampling within their $1\sigma$ intervals.

\begin{figure}[!h]
\begin{center}
\begin{tabular}{cc}
\subfigure[]
{
\begin{minipage}[t]{0.4\linewidth}
\begin{center}
\begin{overpic}[scale=.35]{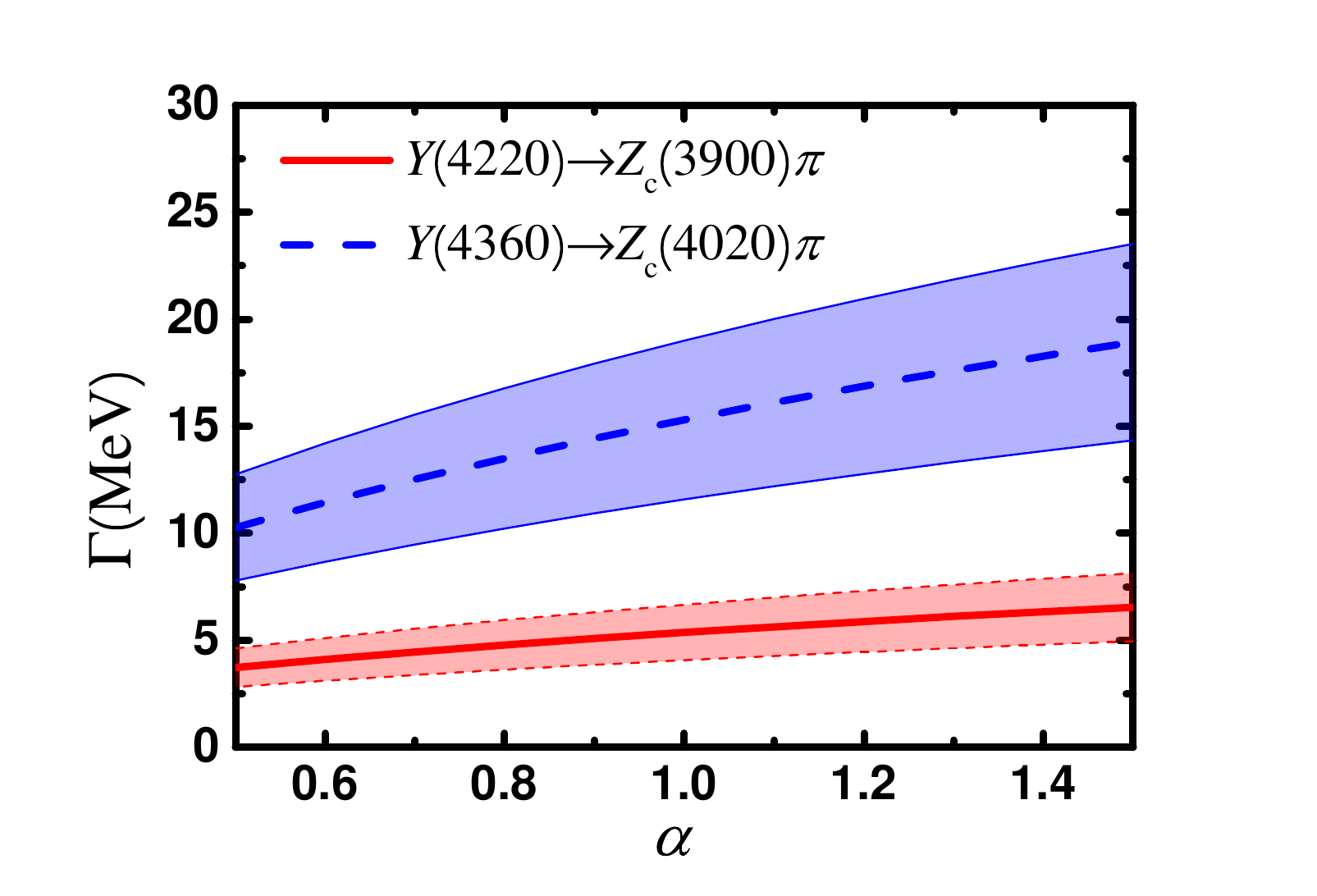}
\end{overpic}
\end{center}
\end{minipage}
} ~~~~~~~~~~
\subfigure[]
{
\begin{minipage}[t]{0.48\linewidth}
\begin{center}
\begin{overpic}[scale=.35]{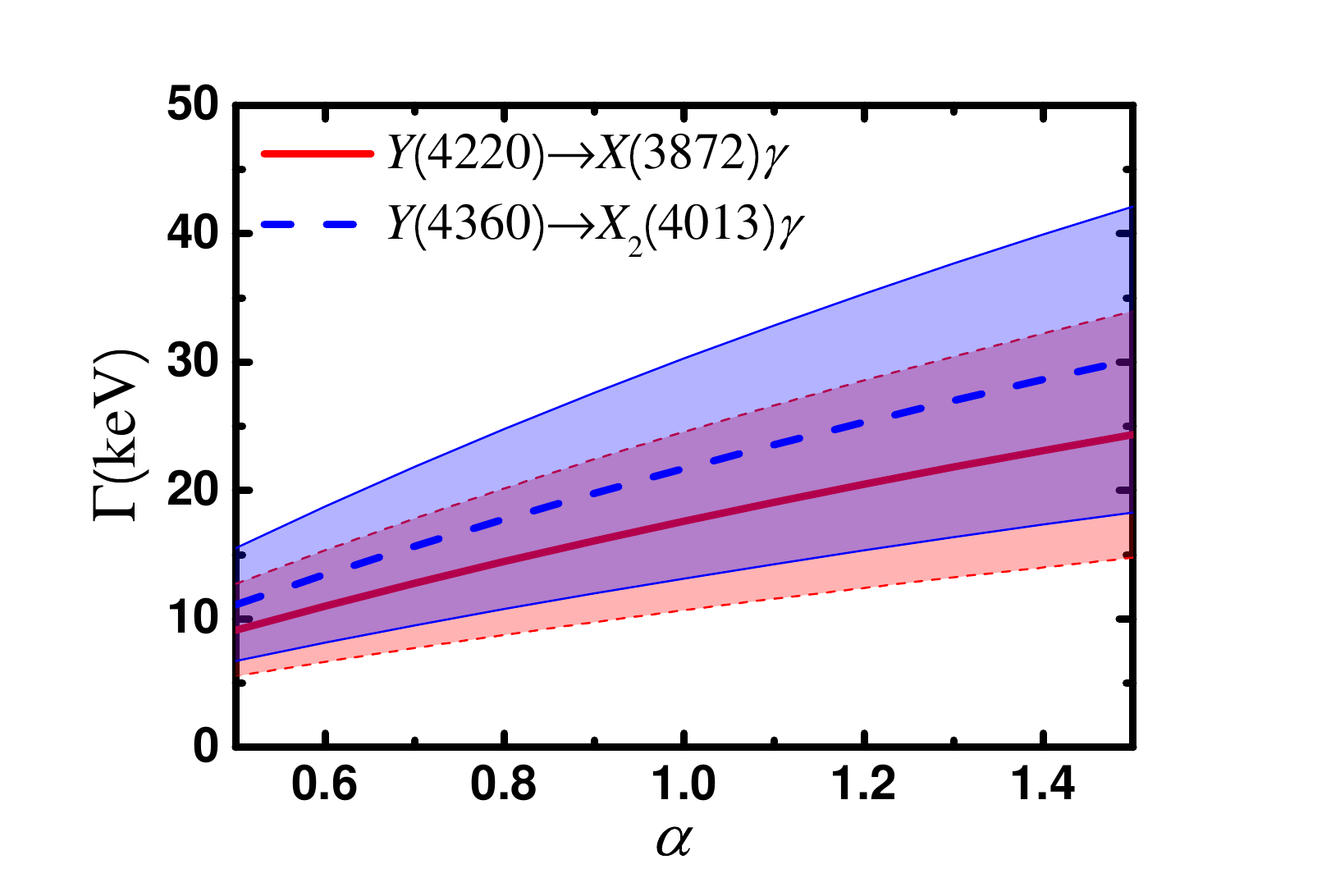}
\end{overpic}
\end{center}
\end{minipage}
}
\end{tabular}
\caption{Partial decay widths of $Y(4220) \to Z_c(3900) \pi $ and $Y(4360) \to Z_c(4020) \pi $ (a),    and  $Y(4220) \to X(3872) \gamma $ and $Y(4360) \to X_2(4013) \gamma $ (b) as a function of $\alpha$. The lines and   bands represent their central values and corresponding uncertainties.       }
\label{ytopi}
\end{center}
\end{figure}

In Fig.~\ref{ytopi}(a), we present the partial decay widths of $Y(4220)\to Z_c(3900)\pi$ and   $Y(4360)\to Z_c(4020)\pi$ as a function of $\alpha$, and their corresponding partial decay widths are in the range of $3.7{ \pm 0.9}\sim6.5{ \pm 1.6}$ MeV and $10.3{ \pm 2.5}\sim 18.9{ \pm 4.6}$ MeV, respectively, in agreement with Refs.~\cite{Dong:2013kta,Chen:2017abq}.    { In Ref.~\cite{Li:2022fgd},   Li et al., estimated the partial decay widths of $Y(4220) \to Z_c(3900) \pi$ and $Y(4360) \to Z_c(4020) \pi$  to be around $1$~MeV and $0.5$~MeV  assuming $YZ$ states as compact tetraquark states,  which are smaller than ours by one order of magnitude.    
}     Considering the widths of $Y(4220)$ and $Y(4360)$, their decaying branching fractions are found to be $0.11{ \pm 0.03}$ and $0.13{ \pm 0.03}$ for $\alpha=1$.   Referring to the RPP, the partial decay width of $Y(4220)  \to J/\psi \pi \pi  $ is around 0.29 MeV and the ratio of the branching fraction $\mathcal{B}[Y(4220) \to Z_c(3900)\pi \to J/\psi \pi \pi ]/\mathcal{B}[Y(4220) \to J/\psi \pi \pi]=0.215$, which imply that the upper limit of the branching fraction $\mathcal{B}[Z_c(3900)\to J/\psi \pi]$ is 0.01 according to our analysis.

For radiative decays,  gauge symmetry must be satisfied. The loop integrals of Eqs.~(\ref{gamma}) can be decomposed into a series of terms in Eq.~(\ref{4220gamma}) and Eq.~(\ref{4360gamma}). In addition, we checked that the loop integrals of Eqs.~(\ref{gamma}) satisfy gauge symmetry, i.e., $p_1 \mathcal{M}_{2a,2b}^{\gamma}=0$ and $p_1 \mathcal{M}_{2c,2d}^{\gamma}=0$.  In Fig.~\ref{ytopi}(b), we present the partial decay widths of $Y(4220) \to X(3872)\gamma$ and $Y(4360) \to X_2(4013)\gamma$ as a function of $\alpha$, and their partial decay widths are $9{ \pm 4}\sim 24{ \pm 10}$ keV and $11{ \pm 4}\sim 30{ \pm 12}$ keV, respectively.  In Ref.~\cite{Dong:2014zka}, Dong et al. estimated the partial decay width of $Y(4220) \to X(3872) \gamma $ to be $23.2\sim 48.6$ keV, a bit larger than ours. We note that the mass of the initial state of this work and that of Ref.~\cite{Dong:2014zka} is a bit different.        
Considering the widths of $Y(4220)$ and $Y(4360)$, we obtain the decay  branching fractions for $\alpha=1$, i.e.,       
$\mathcal{B}[Y(4220) \to X(3872)\gamma]=(3.67{ \pm 1.45})\times 10^{-3}$ and $\mathcal{B}[Y(4360) \to X_2(4013)\gamma]=(1.89{ \pm 0.75})\times 10^{-4}$, which are smaller than those of the above pionic decays by two orders of magnitude.    { In Ref.~\cite{Bruschini:2019qkl},  assuming $Y(4260)$ as a mixture of a $c\bar{c}$ charmonium and a $\bar{D}D_1$ molecule, the partial decay width of $Y(4260) \to X(3872) \gamma$ is estimated to be $54.6 \pm 1.9$~keV, consistent with ours.   In Ref.~\cite{Gens:2021wyf}, Justin M. Gens et al. estimated the partial decay width of  $Y(4220) \to X(3872) \gamma$ to be $100$ keV in the compact tetraquark picture, larger than ours.        
}

\begin{figure}[ttt]
\begin{center}
\begin{tabular}{cc}
\subfigure[]
{
\begin{minipage}[t]{0.40\linewidth}
\begin{center}
\begin{overpic}[scale=.35]{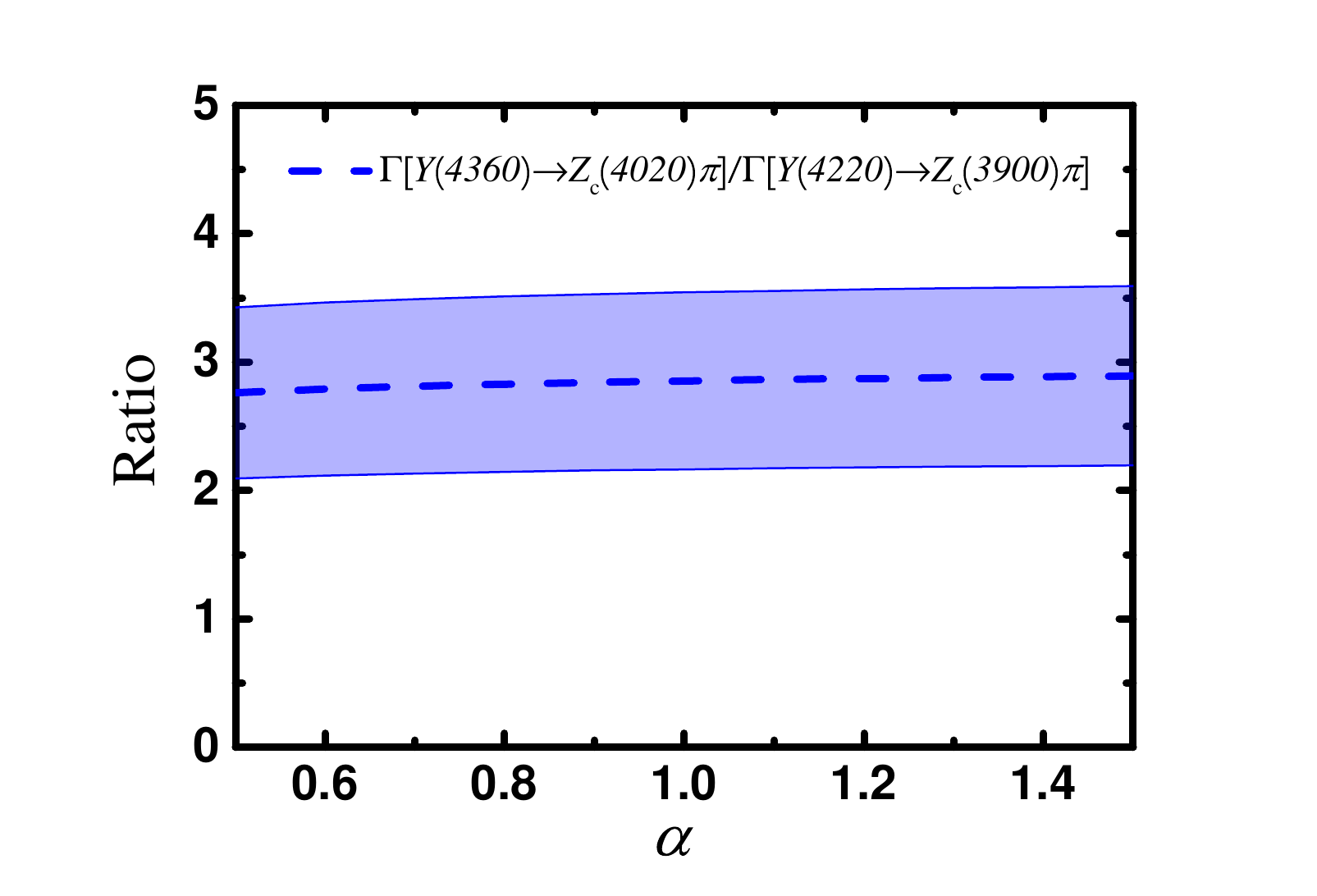}
\end{overpic}
\end{center}
\end{minipage}
} ~~~~~~~~~~~
\subfigure[]
{
\begin{minipage}[t]{0.48\linewidth}
\begin{center}
\begin{overpic}[scale=.35]{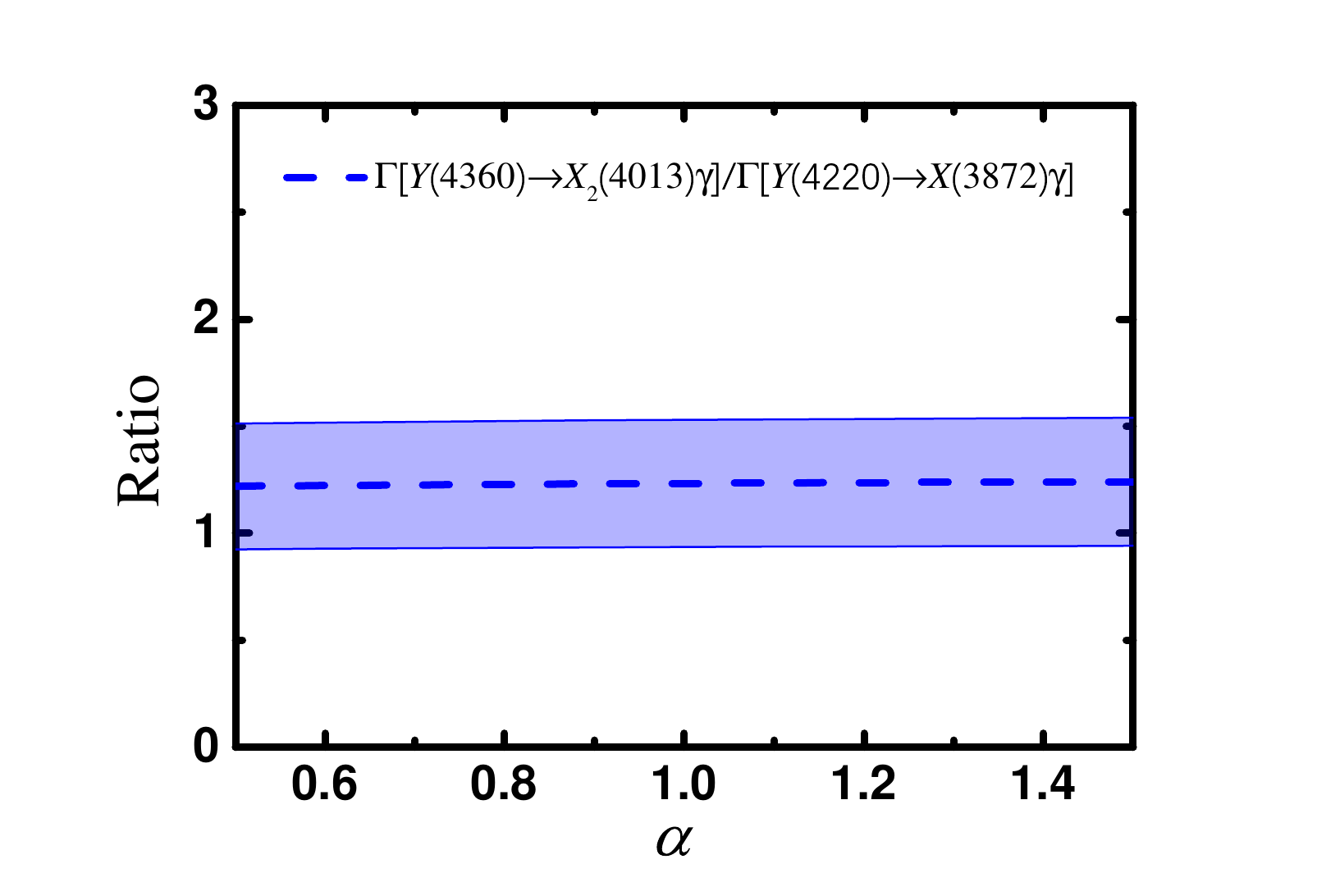}
\end{overpic}
\end{center}
\end{minipage}
}
\end{tabular}
\caption{ Ratios of $\Gamma [Y(4360 ) \to Z_c(4020) \pi ]/\Gamma [Y(4220 ) \to Z_c(3900) \pi ] $ (a) and   $\Gamma [Y(4360 ) \to X_2(4013) \gamma ]/\Gamma [Y(4220  ) \to X(3872) \gamma ] $ (b) as a function of $\alpha$. The lines and bands represent their central values and corresponding uncertainties.    }
\label{ytox}
\end{center}
\end{figure}

The uncertainties in calculating the absolute branching fractions can be reduced for ratios.   
In Fig.~\ref{ytox}(a), we show the ratios of the decay widths of  $Y(4360)\to Z_c(4020)\pi$  to those of  $Y(4220)\to Z_c(3900)\pi$ as a function of $\alpha$. One can see that the result is weakly dependent on the parameter $\alpha$, which is always around  $2.8{ \pm 0.7}$.    {   It is worth noting that  the ratio $\Gamma[Y(4360) \to Z_c(4020) \pi]/\Gamma[Y(4220) \to Z_c(3900) \pi]$ is estimated to be less than but close to $1$  in the compact tetraquark picture~\cite{Li:2022fgd},  smaller than ours by $1.8$.      } Combing the  widths of $Y(4220)$ and $Y(4360)$, we have the ratio $\mathcal{B}[Y(4360) \to Z_c(4020) \pi]/\mathcal{B}[Y(4220) \to Z_c(3900) \pi]=1.2{ \pm 0.3}$, indicating  that the production rate of $Z_c(4020)$ in the $Y(4360)$ decay is a bit larger than that of  $Z_c(3900)$ in the  $Y(4220)$ decay.  Similarly,  the ratio of the partial decay width of  $Y(4360) \to X_2(4013)\gamma$ to that of  $Y(4220) \to X(3872)\gamma$ as a function of $\alpha$ is shown in Fig.~\ref{ytox}(b), which is around $1.2{ \pm 0.3}$. Furthermore, we obtain the ratio   $\mathcal{B}[Y(4360) \to X_2(4013) \gamma]/\mathcal{B}[Y(4220) \to X(3872) \gamma]=0.5{ \pm 0.1}$, implying  that the production rates of $X_2(4013)$ in the $Y(4360)$ decay is lower than that of $X(3872)$ in the $Y(4220)$  decay. The ratios predicted in this work can help verify or refute the molecular interpretation of the $XYZ$ states. 

One should note that the $XYZ$ states are regarded as pure hadronic molecules in this work.  However, recent studies show these $XYZ$ states can not be pure hadronic molecules.   In terms of the mass distribution of $X(3872)$, the $\bar{D}^*D$ molecular component accounts for $82\%$ of the $X(3872)$ total wave function according to the BESIII Collaboration~\cite{BESIII:2023hml}  and  $85\%$ according to the LHCb Collaboration~\cite{LHCb:2020xds}. Very recently,  the $Y(4220)$ is assigned as the $D_1\bar{D}$ hadronic molecule by analyzing its mass distributions in  $e^+e^-$ collisions in the energy region of 4.2 GeV-4.35 GeV~\cite{vonDetten:2024eie}. As for the $Z_c(3900)$, the  $\bar{D}^*D$ molecular component accounts for around $40\%$ of its total wave function by analyzing the BESIII data and the lattice QCD simulations~\cite{Yan:2023bwt}. Given these results, our predictions for the partial decay widths would decrease if other minor components were considered.

\section{Summary and Discussion}
\label{sum}

Since $X(3872)$ was discovered by the Belle Collaboration in 2003, many charmonium-like states beyond the conventional quark model picture (often named $XYZ$ states) have been found experimentally, most of which are close to the mass threshold of a pair of charmed mesons. Recent studies have shown that these  $XYZ$ states have strong couplings to a pair of charmed mesons to the extent that they can be regarded as hadronic molecules. In this work, assuming $X(3872)$, $Z_c(3900)$, and $Y(4220)$ as the $\bar{D}^*D$, $\bar{D}^*D $, and $\bar{D}_1D$ hadronic molecules, we predicted their HQSS partners in the contact range EFT, likely corresponding to $X_2(4013)$, $Z_c(4020)$, and $Y(4360)$. We then employed the triangle diagram mechanism to describe the productions of the $X$ and $Z_c$ states in the radiative and pionic decays of the $Y$ states. Finally, we adopted the effective Lagrangian approach to calculate the partial decay widths.

 Our results showed that the  branching fractions are $\mathcal{B}[Y(4220) \to Z_c(3900) \pi]=0.11{\pm 0.03}$, $\mathcal{B}[Y(4220) \to X(3872)\gamma]=(3.67{ \pm 1.45})\times 10^{-3}$, $\mathcal{B}[Y(4360) \to Z_c(4020) \pi]=0.13{ \pm 0.03}$,  and $\mathcal{B}[Y(4360) \to X_2(4013)\gamma]=(1.89{ \pm 0.75})\times 10^{-4}$, which are dependent on the unknown parameters $\alpha$ in the form factor.     
 In particular, we showed the ratios of the branching fractions $\mathcal{B}[Y(4360) \to Z_c(4020) \pi]/\mathcal{B}[Y(4220) \to Z_c(3900) \pi]=1.2{ \pm 0.3}$ and $\mathcal{B}[Y(4360) \to X_2(4013) \gamma]/\mathcal{B}[Y(4220) \to X(3872) \gamma]=0.5{ \pm 0.1}$, which are  weakly dependent on $\alpha$. From our study, one can conclude that the $Z_c(4020)$ and $X_2(4013)$ can be produced in the pionic and radiative decays of $Y(4360)$, which will serve as a model-independent verification of the molecular nature of the $XYZ$ states if observed.    
 Our studies demonstrated that studying the radiative and pionic decays of the HQSS doublets can deepen one's understanding of the nature of the $ XYZ$ states.

\section{Acknowledgments}
 
This work is partly supported by the National Key R\&D Program of China under Grant No. 2023YFA1606703. M.Z.L
acknowledges support from the National Natural Science Foundation of China under Grant No.12105007.
X.Z.L acknowledges support from the National Natural Science Foundation of China under Grant No.
12247159 and China Postdoctoral Science Foundation under Grant No. 2022M723149.

\appendix

\bibliography{reference}

\end{document}